\documentclass{article}

\usepackage{microtype}
\usepackage[utf8]{inputenc}
\usepackage[T1]{fontenc}
\usepackage{graphicx}
\usepackage{subcaption}
\usepackage{booktabs}
\usepackage{hyperref}

\usepackage[accepted]{icml2026}

\usepackage{amsmath,amssymb,amsthm}
\usepackage{mathtools}
\usepackage{multirow}
\usepackage{listings}
\usepackage{tikz}
\usetikzlibrary{shapes,arrows,positioning,fit,backgrounds,calc,fadings}
\usepackage{fancybox}
\usepackage{framed}

\definecolor{torchslaPurpleStart}{HTML}{4B168C}
\definecolor{torchslaPurpleEnd}{HTML}{7A2CB1}
\definecolor{torchslaOrangeStart}{HTML}{FF9800}
\definecolor{torchslaOrangeEnd}{HTML}{F04B00}
\newcommand{\torchsla}{\texttt{torch-sla}}

\newcommand{\R}{\mathbb{R}}
\newcommand{\bA}{\mathbf{A}}
\newcommand{\bB}{\mathbf{B}}
\newcommand{\bx}{\mathbf{x}}
\newcommand{\bb}{\mathbf{b}}
\newcommand{\bu}{\mathbf{u}}
\newcommand{\bF}{\mathbf{F}}
\newcommand{\bH}{\mathbf{H}}
\newcommand{\bJ}{\mathbf{J}}
\newcommand{\blambda}{\boldsymbol{\lambda}}
\newcommand{\btheta}{\boldsymbol{\theta}}
\newcommand{\by}{\mathbf{y}}
\newcommand{\br}{\mathbf{r}}
\newcommand{\bp}{\mathbf{p}}
\newcommand{\bv}{\mathbf{v}}
\newcommand{\cO}{\mathcal{O}}

\newcounter{algocounter}

\icmltitlerunning{torch-sla: Differentiable Sparse Linear Algebra}

\begin{document}

\twocolumn[
    \icmltitle{\texorpdfstring{torch-sla: Differentiable Sparse Linear Algebra with \\
    Adjoint Solvers and Sparse Tensor Parallelism for PyTorch}{torch-sla: Differentiable Sparse Linear Algebra with Adjoint Solvers and Sparse Tensor Parallelism for PyTorch}}

    \icmlsetsymbol{equal}{*}

    \begin{icmlauthorlist}
        \icmlauthor{Mingyuan Chi}{equal,eth}
        \icmlauthor{Shizheng Wen}{equal,eth}
    \end{icmlauthorlist}

    \vskip 0.08in
    \begin{center}
    \normalsize The library is available at \url{https://www.torchsla.com/}
    \end{center}

    \icmlaffiliation{eth}{ETH Zurich, Zurich, Switzerland}

    \icmlcorrespondingauthor{Mingyuan Chi}{walker.chi.000@gmail.com}
    \icmlcorrespondingauthor{Shizheng Wen}{shizheng.wen@sam.math.ethz.ch}

    \icmlkeywords{Differentiable Sparse Linear Algebra, Scientific Machine Learning, PyTorch}

    \vskip 0.3in
]

\printAffiliationsAndNotice{\icmlEqualContribution}

\begin{abstract}
    Differentiable sparse linear algebra is foundational for scientific 
    machine learning, yet PyTorch lacks a unified library for it: 
    \texttt{torch.sparse} provides only low-level kernels and a 
    non-differentiable, CPU-only \texttt{spsolve}, and \texttt{torch.linalg} 
    is dense-only. We present \torchsla{}, an open-source library that 
    fills this gap. It exposes a single autograd-aware API for direct, 
    iterative, nonlinear, and eigenvalue solvers across five 
    interchangeable backends---SciPy and Eigen on CPU, cuDSS, CuPy, and a 
    PyTorch-native iterative solver on GPU---with automatic dispatch by 
    device and problem size. The library further supports batched solves 
    over shared or distinct sparsity patterns and distributed multi-GPU 
    execution via domain decomposition with halo exchange. These capabilities are made scalable 
    by an O(1)-graph adjoint differentiation framework and an autograd-compatible distributed halo-exchange layer.
    \end{abstract}

\section{Introduction}
\label{sec:intro}

Sparse linear systems arise naturally in many scenarios of 
machine learning and scientific computing. For example, graph neural networks operate 
on sparse adjacency matrices~\citep{kipf2017semi,velivckovic2018graph}; 
neural operators on unstructured meshes often discretize local 
operators sparsely~\citep{li2020fourier,brandstetter2022message,wen2025goat,shi2025mesh}; 
and differentiable simulation pipelines require gradients through 
sparse linear, nonlinear, and eigenvalue 
solves~\citep{holl2024phiflow,hu2019difftaichi,blondel2022efficient}. 
These applications also tend to be batched, and a training step may contain 
one sparse system per sample, time step, or mesh.

In the community of JAX~\citep{jax2018github}, \texttt{jax.lax.custom\_linear\_solve} provides an implicit 
differentiation primitive, \texttt{jax.scipy.sparse.linalg} builds 
differentiable iterative solvers on top of it, and downstream libraries 
extend the same idea to nonlinear and fixed-point problems. By contrast, PyTorch has a 
larger research community, but does not provide an analogous 
stack. \texttt{torch.sparse} exposes
sparse matrix--vector and matrix--matrix multiplication (SpMV and
SpMM), but its only sparse solve, \texttt{spsolve}, is non-differentiable
and implemented as a CPU-only SuperLU~\citep{demmel1999supernodal}
wrapper for square
systems. \texttt{torch.linalg} is dense-only. Calling external solvers such as cuDSS or PETSc from PyTorch breaks the autograd graph,
and simply differentiating through a hand-written iterative solver loop builds an 
$\cO(k)$-node computational graph over the iterations. In practice, users are 
left to drop gradients, move the relevant part of the code to JAX, or 
write a custom solver layer for each problem.

Supporting these workloads in PyTorch takes more than a thin 
\texttt{torch.autograd.Function} wrapper around existing solvers. The 
first issue is backend choice. Direct solvers such as cuDSS and SuperLU
are often fastest below $\sim\!10^5$ degrees of freedom (DOF), but their
$\cO(n^{1.5})$ fill-in~\citep{george1973nested} can exhaust GPU memory
above $\sim\!2\!\times\!10^6$ DOF. Iterative methods such as
conjugate gradient (CG)~\citep{hestenes1952methods} and
biconjugate gradient stabilized (BiCGStab)~\citep{vorst1992bicgstab}
have $\cO(\text{nnz})$ memory cost (with $\text{nnz}$ the number of
non-zeros)
and can reach more than $10^8$ DOF, but small problems are dominated by kernel-launch
overhead. Therefore, a useful PyTorch interface should dispatch across 
heterogeneous backends. The second issue is differentiation. Backpropagating through $k$ 
iterations of an iterative solver stores the intermediate vectors and 
creates an $\cO(k)$-node graph. For a 1M-DOF problem with 1000 CG 
iterations, the saved vectors alone occupy roughly 80~GB. The third issue is scalability.
Problems with $10^7$--$10^9$ unknowns require distributed domain 
decomposition, and the halo exchange used in the forward solve must be 
transposed in the backward pass. Existing PyTorch distributed 
primitives do not provide this sparse-operator adjoint directly. 
\begin{table*}[t]
    \centering
    \caption{Differentiable sparse linear-algebra capability matrix.
    JAX provides a layered ecosystem; PyTorch's native stack stops at
    non-differentiable CPU \texttt{spsolve}; \torchsla{} closes the gap
    in a single library. The matrix compares \emph{first-class}
    support; some ``---'' cells are assemblable by hand (e.g.\
    distributed autograd-compatible solves via
    \texttt{shard\_map}+\texttt{custom\_linear\_solve} in JAX) but are
    not provided as a feature.}
    \label{tab:jax_vs_torch}
    \vspace{0.5em}
    \small
    \begin{tabular}{lccc}
    \toprule
    \textbf{Capability} & \textbf{JAX (native + 3rd party)} & \textbf{PyTorch (native)} & \textbf{\torchsla{}} \\
    \midrule
    Differentiable linear-solve primitive & \texttt{custom\_linear\_solve} & ---            & \checkmark \\
    Differentiable iterative solvers      & \texttt{jax.scipy.sparse.linalg}      & ---            & \checkmark \\
    Differentiable direct solvers         & via Lineax / manual               & ---            & \checkmark \\
    GPU sparse direct backend             & exp. \texttt{spsolve} (no autodiff)  & ---            & cuDSS, CuPy \\
    Nonlinear solve with adjoint          & JAXopt                                & ---            & \checkmark \\
    Eigenvalue solve with adjoint         & dense / problem-specific              & ---            & \checkmark \\
    Batched solves (shared pattern)       & via \texttt{vmap}                     & ---            & \checkmark \\
    Batched solves (distinct patterns)    & manual                                & ---            & \texttt{SparseTensorList} \\
    Distributed sparse solvers + autograd & ---                                   & ---            & \checkmark \\
    \bottomrule
    \end{tabular}
\end{table*}
\paragraph{Contributions.} To address these requirements, \torchsla{} is an open-source PyTorch library for differentiable sparse linear algebra. It provides:
\begin{itemize}
    \setlength{\itemsep}{2pt}
    \item \textbf{Unified backend abstraction} (\S\ref{sec:backend}). 
    A single autograd-aware API dispatches across five 
    backends---SciPy, Eigen, cuDSS, CuPy, and a PyTorch-native 
    iterative solver---selected by device and problem size. The 
    backend interface is also extensible: adding libraries such as 
    PETSc~\citep{balay2019petsc}, Trilinos~\citep{trilinos}, 
    or hypre~\citep{falgout2002hypre} requires only implementing a 
    common backend interface.
    
    \item \textbf{Adjoint differentiation framework}
    (\S\ref{sec:adjoint}). We faithfully port the well-established
    implicit-function-theorem (IFT) adjoint (\S\ref{sec:adjoint} cites
    the JAX and FEniCS equivalents) and apply it uniformly across
    linear, nonlinear, and eigenvalue solves and all five backends,
    keeping the autograd graph at $\cO(1)$ nodes and
    $\cO(\text{nnz})$ memory, independent of solver iterations or
    backend.
    
    \item \textbf{Distributed solvers with autograd-compatible halo 
    exchange} (\S\ref{sec:distributed}). Domain decomposition with 
    halo exchange uses transposed communication in the backward pass, 
    so distributed solves remain compatible with end-to-end differentiation.
\end{itemize}
On 2D Poisson benchmarks, \torchsla{} scales to 169M DOF on one H200 and 400M DOF on three H200s, with gradients verified analytically for linear solves and against finite differences for nonlinear and eigenvalue solves. Moreover, an inverse-coefficient learning task is conducted to demonstrate the end-to-end usability of the library.
\begin{figure*}[t]
    \centering
    \begin{tikzpicture}[
        font=\small,
        >=stealth,
        layer/.style={rectangle, draw=#1!65!black, fill=#1!10, rounded corners=3pt,
            minimum height=8mm, align=center, inner sep=4pt, line width=0.45pt},
        layer/.default=gray,
        band/.style={layer=#1, minimum width=116mm, text depth=0pt},
        type/.style={layer=green, minimum width=25mm, text width=25mm,
            align=center, font=\scriptsize\ttfamily},
        backend/.style={layer=purple, minimum width=20mm, font=\scriptsize\ttfamily},
        ext/.style={layer=gray, minimum width=20mm, font=\scriptsize},
        arrow/.style={->, thick, draw=gray!75!black, line cap=round}
    ]
    
    \node[band=blue] (api) at (0, 4.5) {%
      \textbf{User API}\quad
      \texttt{.solve}\,/\,\texttt{.matvec}\,/\,\texttt{.eigsh}\,/\,\texttt{.det}\,/\,\texttt{nonlinear\_solve}};
    
    \node[type] (st)   at (-4.35, 3.15) {Sparse\\Tensor};
    \node[type] (stl)  at (-1.45, 3.15) {Sparse\\TensorList};
    \node[type] (dst)  at ( 1.45, 3.15) {DSparse\\Tensor};
    \node[type] (dstl) at ( 4.35, 3.15) {DSparse\\TensorList};
    
    \node[band=orange] (adj) at (0, 1.8) {%
      \textbf{Adjoint differentiation framework}\quad
      (\texttt{torch.autograd.Function} wrappers; $\cO(1)$ graph; transposed halo in backward)};
    
    \node[band=yellow] (disp) at (0, 0.45) {%
      \textbf{Backend auto-dispatch}\quad
      (device $\to$ size $\to$ symmetry/SPD $\to$ method)};
    
    \node[backend] (scipy)   at (-4.6, -0.95) {scipy};
    \node[backend] (eigen)   at (-2.3, -0.95) {eigen};
    \node[backend] (cudss)   at ( 0.0, -0.95) {cudss};
    \node[backend] (cupy)    at ( 2.3, -0.95) {cupy};
    \node[backend] (pyt)     at ( 4.6, -0.95) {pytorch};
    
    \node[ext] (super) at (-4.6, -2.15) {SuperLU/UMFPACK};
    \node[ext] (eigl)  at (-2.3, -2.15) {Eigen C++};
    \node[ext] (cudssl)at ( 0.0, -2.15) {NVIDIA cuDSS};
    \node[ext] (cupyl) at ( 2.3, -2.15) {CuPy / cuSPARSE};
    \node[ext] (pytl)  at ( 4.6, -2.15) {torch.sparse};
    
    \node[font=\scriptsize\bfseries, text=gray!70!black] at (-3.45, -2.85) {CPU};
    \node[font=\scriptsize\bfseries, text=gray!70!black] at ( 2.75, -2.85) {GPU};
    \draw[dashed, gray!60] (-1.15, -1.55) -- (-1.15, -3.0);
    
    \draw[arrow] (api) -- (st);
    \draw[arrow] (api) -- (stl);
    \draw[arrow] (api) -- (dst);
    \draw[arrow] (api) -- (dstl);
    \draw[arrow] (st)   -- ($(adj.north)+(-4.35,0)$);
    \draw[arrow] (stl)  -- ($(adj.north)+(-1.45,0)$);
    \draw[arrow] (dst)  -- ($(adj.north)+( 1.45,0)$);
    \draw[arrow] (dstl) -- ($(adj.north)+( 4.35,0)$);
    \draw[arrow] (adj) -- (disp);
    \draw[arrow] (disp.south) -- (scipy.north);
    \draw[arrow] (disp.south) -- (eigen.north);
    \draw[arrow] (disp.south) -- (cudss.north);
    \draw[arrow] (disp.south) -- (cupy.north);
    \draw[arrow] (disp.south) -- (pyt.north);
    \draw[arrow] (scipy)  -- (super);
    \draw[arrow] (eigen)  -- (eigl);
    \draw[arrow] (cudss)  -- (cudssl);
    \draw[arrow] (cupy)   -- (cupyl);
    \draw[arrow] (pyt)    -- (pytl);
    
    \end{tikzpicture}
    \caption{\torchsla{} system architecture. User-facing solve calls
    on any of the four typed sparse tensors flow through a single
    adjoint differentiation layer (implemented as
    \texttt{torch.autograd.Function} wrappers) and a unified
    auto-dispatch policy that selects among five interchangeable
    backends. Backends in turn delegate to established CPU and GPU
    libraries; the same path is reused---in transposed form---during
    the backward pass.}
    \label{fig:architecture}
\end{figure*}

\section{Related Work}
\label{sec:related}

\paragraph{The PyTorch sparse ecosystem.} PyTorch's native sparse
support is fragmented and incomplete. \texttt{torch.sparse} provides
COO/CSR storage and basic kernels (sparse matrix--vector and
matrix--matrix multiplication, SpMV/SpMM, plus elementwise
operations); its only sparse solve, \texttt{torch.sparse.spsolve},
is non-differentiable, CPU-only, and restricted to square systems.
On the other hand, \texttt{torch.linalg} is dense-only. Domain libraries such as
\texttt{torch-scatter}, \texttt{torch-sparse}, and PyTorch
Geometric~\citep{fey2019fast} target message-passing primitives for
graph neural networks (GNNs) rather than general-purpose linear algebra.

\paragraph{Differentiable solvers in PyTorch.} Several PyTorch
libraries already provide differentiable solves, each on a different
slice. Theseus~\citep{pineda2022theseus} is a nonlinear least-squares
\emph{optimizer} whose batched sparse solver is an internal component;
CoLA~\citep{potapczynski2023cola}, the closest analogue,
differentiates an \emph{implicit} linear-operator abstraction that
exploits \emph{algebraic} structure (Kronecker, low-rank) rather than a
sparsity pattern, so its matrix-free operators preclude the sparse
\emph{direct} factorizations (cuDSS) and pattern-based preconditioners
(ILU, AMG) we rely on; PhiFlow~\citep{holl2024phiflow} offers
differentiable preconditioned \emph{iterative} solves but is
GPU-iterative-only. Narrower utilities
(\texttt{torchsparsegradutils}, \texttt{torch\_sparse\_solve}, the
Firedrake--PyTorch bridge) and direct external-solver calls (which
break the autograd graph) cover the rest. No prior PyTorch library
spans direct, iterative, nonlinear, and eigenvalue solves with batched
\emph{and} distributed dispatch at once; \torchsla{} is distinguished
by a GPU sparse-\emph{direct} backend (cuDSS) under autograd,
tensor-parallel distribution with a transposed-halo backward pass, and
this unified coverage at $10^8$+ DOF on one GPU
($4\!\times\!10^8$ distributed). Appendix~\ref{app:related} details the
axes that separate these libraries.

\paragraph{Differentiable solvers in JAX.} JAX provides a layered 
stack for differentiable sparse linear algebra. At the primitive 
layer, \texttt{jax.lax.custom\_linear\_solve} wraps arbitrary 
linear solvers and obtains gradients via the implicit function 
theorem. Built on top, 
\texttt{jax.scipy.sparse.linalg.\{cg, bicgstab, gmres\}} provide 
differentiable iterative solvers; notably, the experimental 
\texttt{sparse.linalg.spsolve} is \emph{not} autodiff-compatible, 
illustrating that even in JAX, differentiable direct solvers
require deliberate adjoint wrapping. At the third-party layer,
Lineax~\citep{rader2023lineax} unifies linear solvers behind a
common interface, while JAXopt~\citep{blondel2022efficient} and
Optimistix~\citep{rader2024optimistix} extend implicit differentiation
to nonlinear and fixed-point problems. The JAX counterpart to our
single-library scope is thus the Lineax\,+\,JAXopt\,+\,Optimistix trio
together; \torchsla{} unifies these axes in one PyTorch library
(Table~\ref{tab:jax_vs_torch}).

\paragraph{Established sparse libraries and GPU backends.} Mature 
sparse-solver libraries span CPU and GPU, direct and iterative, 
single-node and distributed regimes. SciPy~\citep{scipy2020} 
provides SuperLU on CPU; NVIDIA 
cuDSS~\citep{nvidia2024cudss} offers GPU direct solvers (LU,
Cholesky, LDLT); CuPy~\citep{okuta2017cupy} exposes iterative and direct solvers on GPU; and AmgX~\citep{naumov2015amgx}
provides algebraic multigrid with strong industrial scalability. At larger scale, 
PETSc~\citep{balay2019petsc}, Trilinos~\citep{trilinos}, 
and hypre~\citep{falgout2002hypre} provide distributed sparse 
linear algebra with mature preconditioners; however, these systems generally do not compose with PyTorch
automatic differentiation, and exposing them through PyTorch typically
requires copying tensors across language, device, or process boundaries.  
\torchsla{} instead treats such solvers as interchangeable,
autograd-aware backends behind a single PyTorch API. New backends only
need to conform to a common solver interface, while the distributed
domain-decomposition patterns used in PETSc, Trilinos, and
OpenFOAM~\citep{jasak2007openfoam} are
adapted to PyTorch with autograd-compatible communication
(\S\ref{sec:distributed}).
\section{Methodology}
\label{sec:method}

\torchsla{} is built on three components: a unified backend
abstraction (\S\ref{sec:backend}) that dispatches across five
solvers behind one autograd-aware API, an adjoint differentiation
framework (\S\ref{sec:adjoint}) that keeps the autograd graph at
$\cO(1)$ nodes regardless of solver iterations, and an
autograd-compatible distributed layer (\S\ref{sec:distributed})
implementing domain decomposition with halo exchange.
Figure~\ref{fig:architecture} gives a system overview.

\subsection{Unified Backend Abstraction}
\label{sec:backend}

\paragraph{Typed sparse-tensor hierarchy.} \torchsla{} exposes four 
sparse-tensor types organized along two orthogonal axes---single 
matrix versus list, local versus distributed:
\begin{center}
\footnotesize
\begin{tabular}{@{}lll@{}}
\toprule
\textbf{Layout}       & \textbf{Single matrix}         & \textbf{Matrix list} \\
\midrule
\textbf{Local}        & \texttt{SparseTensor}          & \texttt{SparseTensorList}              \\
\textbf{Distributed}  & \texttt{DSparseTensor}         & \texttt{DSparseTensorList}             \\
\bottomrule
\end{tabular}
\end{center}
\texttt{SparseTensor} holds a single matrix or a batch sharing one
sparsity pattern, so one symbolic factorization (direct) or set of
communication buffers (iterative) is reused across the batch.
\texttt{SparseTensorList} holds a batch with \emph{distinct} patterns
(GNN minibatches, neural operators on irregular meshes), dispatching
each element with an isolated autograd graph. \texttt{DSparseTensor}
and \texttt{DSparseTensorList} are distributed variants where each
process owns a row partition with halo metadata. All types expose the
same methods (\texttt{.solve}, \texttt{.matvec}, \texttt{.eigsh},
\texttt{.det}) plus conversion utilities (\texttt{partition},
\texttt{gather\_global}, etc.).

\paragraph{Five backends and auto-dispatch.}
\torchsla{} ships with five backends, each specialized to a device and
a problem-size regime: \texttt{scipy}~\citep{scipy2020} (CPU;
SuperLU, UMFPACK~\citep{davis2004umfpack}, CG,
BiCGStab; CPU default with machine precision),
\texttt{eigen}~\citep{eigenweb} (CPU;
CG, BiCGStab; alternative CPU iterative), \texttt{cudss} (CUDA; LU,
Cholesky, LDLT; fastest direct solver below $\sim\!2$M DOF),
\texttt{cupy}~\citep{okuta2017cupy} (CUDA; LU, CG, and generalized
minimal residual (GMRES) via \texttt{cupyx}), and a
\texttt{pytorch}-native backend (CUDA; CG, BiCGStab; beyond $\sim\!2$M DOF on a single GPU).
We list the backend library versions used in our experiments in
Appendix~\ref{app:backends}; full method-level coverage per backend is
also given there.

The auto-dispatch policy follows three rules in priority order:
(i) match the device of the input tensors; (ii) for CUDA devices,
prefer \texttt{cudss} when the matrix fits within the fill-in budget
($n \lesssim 2\!\times\!10^6$ for typical 2D/3D PDE matrices) and
fall back to the \texttt{pytorch}-native iterative backend for
larger problems; (iii) for CPU devices, prefer \texttt{scipy} for
direct solves and offer \texttt{eigen} as an iterative alternative.
Symmetry and symmetric positive-definiteness (SPD) are detected on
the matrix values and used to upgrade general LU to Cholesky or LDLT
where applicable. Adding a new backend---PETSc, Trilinos, hypre, or a
learned preconditioner---requires only implementing the solver
methods exposed by the backend and registering its applicability
conditions through \texttt{select\_backend}.

\paragraph{Bridging backends into autograd.} All five backends are
wrapped uniformly in \texttt{torch.autograd.Function}: \texttt{forward}
invokes the backend and stashes $(\bA,\bx)$, \texttt{backward} runs the
adjoint solve of \S\ref{sec:adjoint}. For the external backends this is
the only path preserving the autograd graph across the C++/CUDA
boundary; the PyTorch-native backend (otherwise an $\cO(k)$-node graph)
is wrapped the same way under \texttt{torch.no\_grad()}, so the adjoint
framework is the single gradient path for all backends.

\paragraph{API.} A single \texttt{.solve(b)} call dispatches uniformly
across all five backends and four tensor types---single, batched
(shared or distinct patterns), nonlinear, and distributed
solves---with gradients flowing through unchanged; \texttt{backend} and
\texttt{method} keywords override auto-dispatch.
Appendix~\ref{app:api} gives end-to-end code.

\subsection{Adjoint Differentiation Framework}
\label{sec:adjoint}

The adjoint differentiation in \torchsla{} faithfully ports a
classical, well-understood technique---the implicit-function-theorem
(IFT) adjoint underlying
\texttt{jax.lax.custom\_linear\_solve}~\citep{jax2018github},
JAXopt~\citep{blondel2022efficient},
Optimistix~\citep{rader2024optimistix},
CoLA~\citep{potapczynski2023cola}, and the dolfin-adjoint/pyadjoint
tape~\citep{mitusch2019dolfinadjoint}. Our contribution is its
uniform, backend-agnostic integration into PyTorch autograd at scale,
not the derivation; we restate it here to fix notation.

Naively backpropagating through $k$ iterations of an iterative
solver builds a graph with $\cO(k)$ nodes and stores intermediate
vectors at each step; for a 1M-DOF problem with 1000 CG iterations,
the saved vectors alone occupy roughly $80$~GB. The adjoint
framework collapses this to $\cO(1)$ nodes and
$\cO(n + \text{nnz})$ memory, independent of $k$ and of the
backend used for the forward solve.

\subsubsection{The General Setting}
\label{sec:adjoint_general}

Let $\bx^*(\btheta) \in \R^n$ be the solution of an implicit residual 
equation
\begin{equation}
    \bF(\bx^*, \btheta) \;=\; \mathbf{0}, 
    \qquad \bF: \R^n \times \R^d \to \R^n,
    \label{eq:residual}
\end{equation}
where $\btheta \in \R^d$ collects all differentiable inputs---matrix
non-zero values, right-hand side entries, or physical parameters.
For a downstream scalar loss $\mathcal{L}(\bx^*)$, the gradient follows
from the method of Lagrange multipliers: stationarity of the augmented
objective
$\mathcal{A} = \mathcal{L}(\bx) + \blambda^\top \bF(\bx,\btheta)$
in $\bx$ and $\blambda$ identifies $\blambda$ as the multiplier
enforcing the constraint and gives, equivalently to the implicit
function theorem~\citep{krantz2002implicit},
\begin{equation}
\begin{aligned}
    \frac{\partial \mathcal{L}}{\partial \btheta}
    &= -\,\blambda^\top \frac{\partial \bF}{\partial \btheta},
    &
    \bJ^\top \blambda
    &= \frac{\partial \mathcal{L}}{\partial \bx^*}, \\
    \bJ
    &\equiv \frac{\partial \bF}{\partial \bx^*}\bigg|_{\bx^*}.
\end{aligned}
    \label{eq:adjoint_master}
\end{equation}
The backward pass thus reduces to one adjoint solve
$\bJ^\top \blambda = \partial \mathcal{L}/\partial \bx^*$ at the
converged solution, followed by a vector--Jacobian product
$-\blambda^\top \partial \bF / \partial \btheta$. Only $\bx^*$ and
the data needed to apply $\bJ$ are stashed during the forward pass;
intermediate solver iterates are not referenced.

\subsubsection{Three Instances}
\label{sec:adjoint_instances}

We instantiate Eq.~\eqref{eq:adjoint_master} for the three solver
types in \torchsla{}.

\paragraph{Linear systems.} With residual 
$\bF(\bx, \bA, \bb) = \bA\bx - \bb$, the Jacobian is $\bJ = \bA$ 
and Eq.~\eqref{eq:adjoint_master} specializes to the single 
adjoint linear system $\bA^\top \blambda = \partial\mathcal{L}/\partial\bx$, 
yielding
\begin{equation}
    \frac{\partial \mathcal{L}}{\partial \bb} \;=\; \blambda, 
    \qquad
    \frac{\partial \mathcal{L}}{\partial \bA_{ij}} \;=\; -\,\blambda_i\, x_j.
    \label{eq:adjoint_linear}
\end{equation}
The matrix gradient is materialized only on the sparsity pattern at 
total cost $\cO(\text{nnz})$.

\paragraph{Nonlinear systems.} For a general residual 
$\bF(\bu, \btheta) = \mathbf{0}$ converged by Newton, Picard, or 
Anderson acceleration~\citep{anderson1965iterative,kelley1995iterative} 
to a fixed point $\bu^*$, Eq.~\eqref{eq:adjoint_master} applies 
directly with $\bJ = \partial \bF / \partial \bu^*$ evaluated at the 
solution. The forward pass may take many nonlinear iterations, each 
itself involving a linear solve; the backward pass is one adjoint 
linear solve plus one vector--Jacobian product. The
Jacobian-vector and vector-Jacobian products required to apply $\bJ$ 
and $\bJ^\top$ are obtained from PyTorch's autograd via 
\texttt{torch.autograd.functional.\{jvp,vjp\}}, so users supplying
a nonlinear residual as a Python function obtain a matrix-free
adjoint without writing additional code. Because the adjoint is taken
at the converged state, it is exact only once
$\bF(\bu^*,\btheta)\approx\mathbf{0}$; early termination biases the
gradient. The $\cO(\text{nnz})$ matrix-gradient assembly of
Eq.~\eqref{eq:adjoint_linear} carries over to this non-affine case.

\paragraph{Eigenvalue problems.} For the symmetric eigenvalue 
problem $\bA \bv = \lambda \bv$ with $\|\bv\|_2 = 1$, the 
Hellmann--Feynman theorem~\citep{magnus1985differentiating} gives 
the closed form
\begin{equation}
    \frac{\partial \lambda}{\partial \bA_{ij}} \;=\; v_i\, v_j,
    \label{eq:adjoint_eig}
\end{equation}
which is Eq.~\eqref{eq:adjoint_master} specialized to the 
constrained residual 
$\bF(\bv, \lambda; \bA) = \big(\bA \bv - \lambda \bv,\; 
\tfrac{1}{2}(\bv^\top \bv - 1)\big)$ after eliminating the Lagrange 
multiplier on the normalization. Eigenvector gradients require one 
additional deflated linear solve per eigenpair; eigenvalue
gradients reduce to an outer product on the sparsity pattern at
cost $\cO(\text{nnz})$. Eq.~\eqref{eq:adjoint_eig} assumes a
\emph{simple} eigenvalue with a smoothly varying eigenvector; at
crossings or clusters the eigenvector gradient is ill-defined and needs
degenerate-perturbation handling, so \torchsla{} targets the simple
case (\S\ref{sec:conclusion}).

\subsubsection{Complexity}
\label{sec:adjoint_complexity}

Table~\ref{tab:complexity} summarizes the savings for the linear 
case; the nonlinear and eigenvalue cases inherit the same $\cO(1)$ 
graph-node count.

\begin{table}[h]
    \centering
    \caption{Complexity of naive autograd-through-iterations versus 
    \torchsla{}'s adjoint backward, for a sparse linear solve with $n$ 
    unknowns, $\text{nnz}$ non-zero entries, $k$ solver iterations, and 
    forward solve time $T_{\text{solve}}$.}
    \label{tab:complexity}
    \vspace{0.5em}
    \small
    \begin{tabular}{lcc}
    \toprule
                                  & \textbf{Naive} & \textbf{Adjoint} \\
    \midrule
    Graph nodes                   & $\cO(k)$       & $\cO(1)$ \\
    Graph memory                  & $\cO(k\, n)$   & $\cO(n + \text{nnz})$ \\
    Backward time                 & $\cO(k \cdot \text{nnz})$ & $\cO(T_{\text{solve}}) + \cO(\text{nnz})$ \\
    \bottomrule
    \end{tabular}
\end{table}

The forward-pass cost is unchanged at $T_{\text{solve}}$. The 
backward pass executes (i) one adjoint solve of the same size and 
sparsity pattern as the forward---reusing the same backend and, 
where applicable, the same factorization---and (ii) an 
$\cO(\text{nnz})$ outer-product evaluation to assemble the matrix 
gradient. Since Eq.~\eqref{eq:adjoint_master} treats the forward
solve as a black box, any of the five backends of
\S\ref{sec:backend} may be used for the forward solve, and the
adjoint solve may even use a different backend. Composition with
the distributed layer of \S\ref{sec:distributed} is via the
transposed halo exchange formalized in
Appendix~\ref{app:dist_cg}.

\subsection{Distributed Layer with Halo Exchange}
\label{sec:distributed}

For problems beyond single-device memory, \torchsla{} partitions
$\bA$ across processes following the standard
PETSc/Trilinos/OpenFOAM pattern: row-block ownership with halo
metadata, halo exchange before each SpMV, and \texttt{all\_reduce}
for global inner products. Each forward halo exchange induces a
\emph{transposed} halo exchange in the backward pass, and the
distributed adjoint solve is itself an instance of
Eq.~\eqref{eq:adjoint_master} on the distributed residual.

\paragraph{Domain decomposition.} A sparse matrix
$\bA \in \R^{n \times n}$ is partitioned across $P$ processes.
Process $p$ owns a contiguous block of rows
$\mathcal{O}_p \subset \{1,\ldots,n\}$ and stores
$\bA[\mathcal{O}_p, :]$ together with a halo index set
$\mathcal{H}_p$ enumerating the column indices outside
$\mathcal{O}_p$ that appear in any locally owned row. Before each
local SpMV, processes exchange boundary-owned entries with
neighbors so that halo values are current; the local SpMV
$\by_{\mathcal{O}_p} \gets \bA[\mathcal{O}_p,:]\,\bx_{\text{local}}$
is then purely local. \torchsla{} supports contiguous
row-based partitioning, recursive coordinate
bisection~\citep{berger1987partitioning} when node
coordinates are available, and edge-cut minimization via
METIS~\citep{karypis1998fast} for unstructured meshes through
\texttt{partition\_simple} and related utilities.

\paragraph{Distributed Krylov solvers.} Distributed CG performs
one halo exchange per iteration (inside the SpMV) plus two
\texttt{all\_reduce} operations for the inner products
$\langle \br, \br \rangle$ and $\langle \bp, \bA\bp \rangle$,
giving per-iteration cost
$\cO(|\mathcal{H}_p| + \log P)$ with
$|\mathcal{H}_p| \sim \cO((n/P)^{(d-1)/d})$ on $d$-dimensional
meshes with balanced partitioning. All tensor operations remain
on-device, so the NVIDIA Collective Communications Library
(NCCL)~\citep{nvidia_nccl} backend moves data directly between GPU
memories without host staging. BiCGStab and the locally optimal block
preconditioned conjugate gradient (LOBPCG)
eigensolver~\citep{knyazev2001lobpcg} follow the
same template, substituting the Krylov recurrence and adding
global reductions for additional inner products as needed.
Pseudocode is in Appendix~\ref{app:dist_cg}.

\paragraph{Autograd composition: the transposed halo exchange.}
Treating the halo exchange as a linear operator
$\bH: \R^{|\mathcal{O}_p|} \to \R^{|\mathcal{O}_p| + |\mathcal{H}_p|}$
that scatters owned values into the halo positions of neighbor
processes, the forward distributed SpMV factors as
\begin{equation}
    \by_{\mathcal{O}_p} \;=\; \bA_{\text{local}}\,\bH(\bx_{\mathcal{O}_p}),
    \label{eq:forward_spmv}
\end{equation}
with adjoint
\begin{equation}
    \frac{\partial \mathcal{L}}{\partial \bx_{\mathcal{O}_p}}
    \;=\;
    \bH^\top\!\left(\bA_{\text{local}}^\top\,
                    \frac{\partial \mathcal{L}}{\partial \by_{\mathcal{O}_p}}\right).
    \label{eq:backward_spmv}
\end{equation}
The transposed halo exchange $\bH^\top$ uses the same neighbor
graph and message sizes as $\bH$, with reversed sender/receiver
roles and summation rather than overwrite at the receive site:
where $\bH$ sends owned boundary values to neighbor halos,
$\bH^\top$ sums incoming halo gradients back into the owning
process's boundary entries. \torchsla{} implements both $\bH$ and
$\bH^\top$ as PyTorch \texttt{autograd.Function}s on top of
\texttt{torch.distributed} primitives, so each call to distributed
SpMV inside CG, BiCGStab, or LOBPCG generates the correct
backward communication automatically. Composing this with
\S\ref{sec:adjoint}, a distributed linear solve runs distributed
CG forward and one adjoint distributed solve
$\bA^\top \blambda = \partial\mathcal{L}/\partial\bx$ backward,
reusing the same decomposition, backend, and neighbor graph;
matrix gradients $-\lambda_i x_j$ are assembled locally on each
process's owned non-zeros, with no additional communication.

\paragraph{Scope of distributed gradients.} \texttt{matvec},
\texttt{solve}, and \texttt{eigsh} support distributed gradient
flow with NCCL on GPU and Gloo on CPU. \texttt{det} is a global scalar
that cannot be computed without full matrix information; the
distributed implementation gathers all partitions onto one rank and
emits a runtime warning, and is documented as not scaling to
distributed sizes (a distributed factorization or stochastic
log-determinant is left to future work).

\section{Experiments}
\label{sec:experiments}

We evaluate \torchsla{} along four axes corresponding to the 
three components of \S\ref{sec:method} plus an end-to-end demonstration: scalability across single- and
multi-GPU regimes (\S\ref{sec:exp_scalability}), the $\cO(1)$
autograd-graph claim of the adjoint framework
(\S\ref{sec:exp_adjoint}), gradient verification
(\S\ref{sec:exp_gradient}), and end-to-end usability through an
inverse coefficient-learning task (\S\ref{sec:exp_inverse}).

\subsection{Scalability}
\label{sec:exp_scalability}

\begin{table}[t]
    \centering
    \caption{Single-GPU benchmark on 2D Poisson, H200 GPU,
    \texttt{float64}. SciPy and cuDSS are direct solvers; the
    \textbf{CG} column is \torchsla{}'s \texttt{pytorch}-native
    iterative backend (Jacobi-preconditioned), with its peak memory
    (\textbf{Mem.}) and final residual (\textbf{Resid.}).}
    \label{tab:benchmark}
    \vspace{0.5em}
    \scriptsize
    \setlength{\tabcolsep}{4pt}
    \begin{tabular}{rrrrrr}
    \toprule
    \textbf{DOF} & \textbf{SciPy} & \textbf{cuDSS} & \textbf{CG} & \textbf{Mem.} & \textbf{Resid.} \\
    \midrule
    10K   & 24 ms   & 128 ms  & 20 ms   & 36 MB    & $10^{-9}$ \\
    100K  & 29 ms   & 630 ms  & 43 ms   & 76 MB    & $10^{-7}$ \\
    1M    & 19.4 s  & 7.3 s   & 190 ms  & 474 MB   & $10^{-7}$ \\
    2M    & 52.9 s  & 15.6 s  & 418 ms  & 916 MB   & $10^{-7}$ \\
    16M   & OOM     & OOM     & 7.3 s   & 7.1 GB   & $10^{-6}$ \\
    169M  & OOM     & OOM     & 224 s   & 74.8 GB  & $10^{-6}$ \\
    \bottomrule
    \end{tabular}
\end{table}
\paragraph{Single-GPU scalability.}

Table~\ref{tab:benchmark} compares solver backends across problem sizes spanning five orders of 
magnitude. Three regimes emerge along the scale axis. Below 100K DOF, direct
solvers dominate: SciPy SuperLU reaches machine precision
($10^{-14}$) in 24~ms, while GPU launch overhead leaves cuDSS at
128~ms. Between 100K and 2M DOF, iterative solvers become
competitive in runtime, while direct solvers remain valuable when
near-machine-precision accuracy is required. At 1M DOF, the
\torchsla{} CG backend (190~ms) outperforms cuDSS (7.3~s) by $38\times$.
Above 2M DOF, direct solvers run out of memory while CG's
near-linear scaling extends to 169M DOF within 74.8~GB. Fitting
$T = c \cdot n^\alpha$ to the CG measurements gives
$\alpha \approx 1.1$, consistent with the theoretical
$\cO(\sqrt{\kappa} \cdot \text{nnz})$ iteration cost and condition
number $\kappa \sim n$ for 2D
Poisson under Jacobi preconditioning~\citep{trefethen1997numerical};
the measured 443 bytes/DOF is roughly $3\times$ a minimal-storage
estimate, broken down in Appendix~\ref{app:backends}.

\paragraph{Multi-GPU scaling.}
Table~\ref{tab:distributed}
summarizes distributed CG performance on multiple H200 GPUs with the
NCCL backend. This experiment demonstrates \emph{memory capacity and
per-iteration throughput} of the distributed forward/backward path,
not solver convergence: under a fixed 1000-iteration budget with only
Jacobi preconditioning, the residual at $10^8$--$4\!\times\!10^8$ DOF
stays in the $10^{-2}$ range, far from a converged tolerance. Reaching
a meaningful tolerance at this scale needs a stronger preconditioner
(e.g.\ algebraic multigrid via AmgX/hypre), which we leave to future
work (\S\ref{sec:conclusion}).
\begin{table}[t]
\centering
\caption{Distributed CG on H200 GPUs with NCCL, fixed 1000-iteration
budget (Jacobi-preconditioned). The residual (\textbf{Resid.}) is the
state after the budget, not convergence (see text and
\S\ref{sec:conclusion}).}
\label{tab:distributed}
\vspace{0.5em}
\scriptsize
\setlength{\tabcolsep}{4pt}
\begin{tabular}{rrrrr}
\toprule
\textbf{DOF} & \textbf{GPUs} & \textbf{Time} & \textbf{Mem./GPU} & \textbf{Resid.} \\
\midrule
100M  & 4 & 36.1 s   & 23.3 GB  & $1.0 \times 10^{-2}$ \\
200M  & 3 & 120 s    & 53.7 GB  & $1.5 \times 10^{-2}$ \\
300M  & 3 & 217 s    & 80.5 GB  & $1.9 \times 10^{-2}$ \\
400M  & 3 & 331 s    & 110 GB   & $2.3 \times 10^{-2}$ \\
\bottomrule
\end{tabular}
\end{table}
In memory and throughput, \torchsla{} scales to 400M DOF on 3 H200
GPUs. From 1M to 100M DOF time scales near-linearly
($T \propto n^{1.05}$), reaching 2.8M DOF/s---near the aggregate
bandwidth of four H200s. Above 100M DOF we use 3 GPUs (a
node-allocation constraint, not memory: per-GPU peak is 110~GB of the
140~GB H200), sustaining 1.2M DOF/s. Per-GPU memory reaches 275~B/DOF
and scales as $\cO(n/P + |\mathcal{H}_p|)$,
$|\mathcal{H}_p| \sim \cO(\sqrt{n/P})$ on 2D grids, as predicted in
\S\ref{sec:distributed}.

\subsection{Adjoint vs.\ Naive Backpropagation}
\label{sec:exp_adjoint}

This experiment directly tests the central claim of
\S\ref{sec:adjoint}: that the adjoint backward pass produces an
$\cO(1)$ autograd graph regardless of forward iteration count,
whereas naive backpropagation through CG iterations produces an
$\cO(k)$-node graph that grows linearly in memory and time. We
solve the same 2D Poisson problem with two paths through the same
\texttt{pytorch}-native CG forward kernel:
\textbf{naive}, a manual CG implemented entirely in autograd-tracked
PyTorch ops so that every iteration adds nodes to the graph, and
\textbf{adjoint}, the default \torchsla{} path which wraps the
solve in a \texttt{torch.autograd.Function} and computes gradients
via Eq.~\eqref{eq:adjoint_master}. Both paths use vanilla
unpreconditioned CG forced to run exactly $k$ iterations, and we
sweep $k \in \{10, 50, 100, 200, 500, 1000, 2000, 5000\}$. To
isolate the autograd-graph cost from the SpMV implementation, the
naive baseline uses a hand-coded scatter-based SpMV
(\texttt{val} $\cdot$ \texttt{x[col]} followed by
\texttt{index\_add}); the otherwise tempting
\texttt{torch.sparse.mm} produces a \emph{dense} backward gradient
with respect to the matrix values and OOMs at $k\!=\!1$ on this
problem size. In this experiment, we use a single NVIDIA RTX PRO 6000 Blackwell (96~GB).

\begin{figure}[t]
\centering
\includegraphics[width=0.95\linewidth]{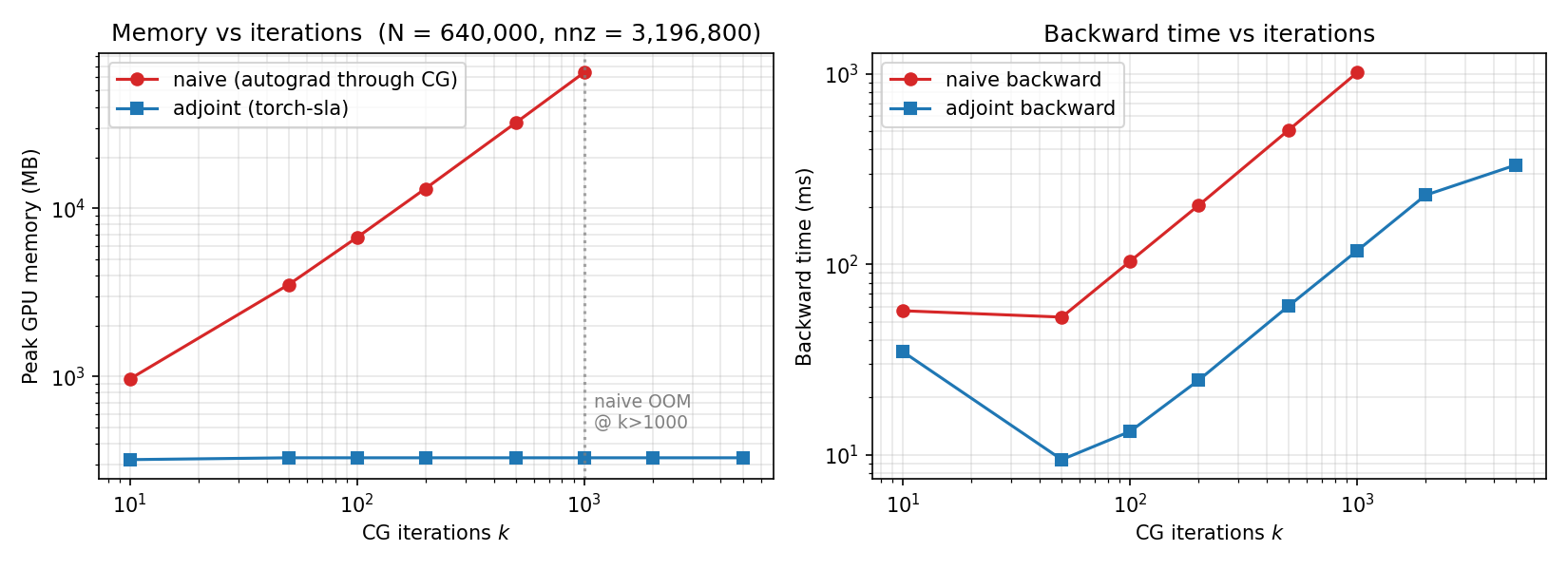}
\caption{Adjoint vs.\ naive backprop through $k$ CG iterations
(2D Poisson, $N = 640{,}000$; RTX PRO 6000, \texttt{float64}).
\textbf{Left}: adjoint memory is flat ($\sim\!330$~MB); naive grows
linearly and OOMs at $k \geq 2000$. \textbf{Right}: naive backward time
grows linearly in $k$, while the adjoint path is dominated by the
constant backward-solve cost $T_{\text{solve}}$ (its mild 35--117~ms
growth over $k=10$--1000 is that solve run to the same $k$, not a
$k$-dependent $\cO(1)$ graph; cf.\ Table~\ref{tab:complexity}).}
\label{fig:adjoint_vs_naive}
\end{figure}

\paragraph{Results.} The naive path materializes 
$\sim\!64$~MB of autograd-tracked intermediates per CG iteration---two $\text{nnz}$-sized tensors from 
the scatter-based SpMV ($\sim\!51$~MB at $\text{nnz} = 3.2\times 10^6$) 
plus a handful of $N$-vectors from the Krylov recurrence 
($\sim\!13$~MB)---growing to $64.1$~GB at $k\!=\!1000$ and OOM-ing at $k\!=\!2000$ on 
the 96~GB device, where it would need $\approx\!128$~GB of 
intermediates. The adjoint path holds steady at $\sim\!328$~MB 
across the entire sweep---a near-perfect slope-one line for naive 
versus a flat line for adjoint on the log-log left panel of 
Figure~\ref{fig:adjoint_vs_naive}---in agreement with the 
$\cO(n + \text{nnz})$ floor of Table~\ref{tab:complexity}. At 
$k\!=\!1000$ this is a $195\times$ memory reduction. On the time 
axis (right panel), naive backward grows linearly with $k$ at 
roughly $1$~ms per iteration and is $4\!-\!9\times$ slower than 
adjoint across the regime where naive runs at all 
($k = 100\ldots 1000$). 
On a smaller version of the same 2D Poisson problem as in 
\S\ref{sec:exp_adjoint} (identical five-point stencil and 
homogeneous Dirichlet boundary convention, $n_{\text{grid}} = 64$, 
$N = n_{\text{grid}}^2 = 4096$ interior unknowns), where both paths 
can be run to full convergence (atol $= 10^{-12}$, $k = 3000$), the 
loss values agree to machine precision ($1.96 \times 10^{-16}$ 
relative error)and to $10^{-14}\!-\!10^{-4}$ on gradients with respect to $\bb$ and 
$\bA$. Together, these results validate both halves of the 
central claim of \S\ref{sec:adjoint}: the autograd graph is 
$\cO(1)$ in $k$, and the resulting gradients are correct. 
Appendix~\ref{app:adjoint_extra} reports the per-$k$ computational cost and correctness 
analysis.

\begin{figure*}[t]
\centering
\includegraphics[width=\linewidth]{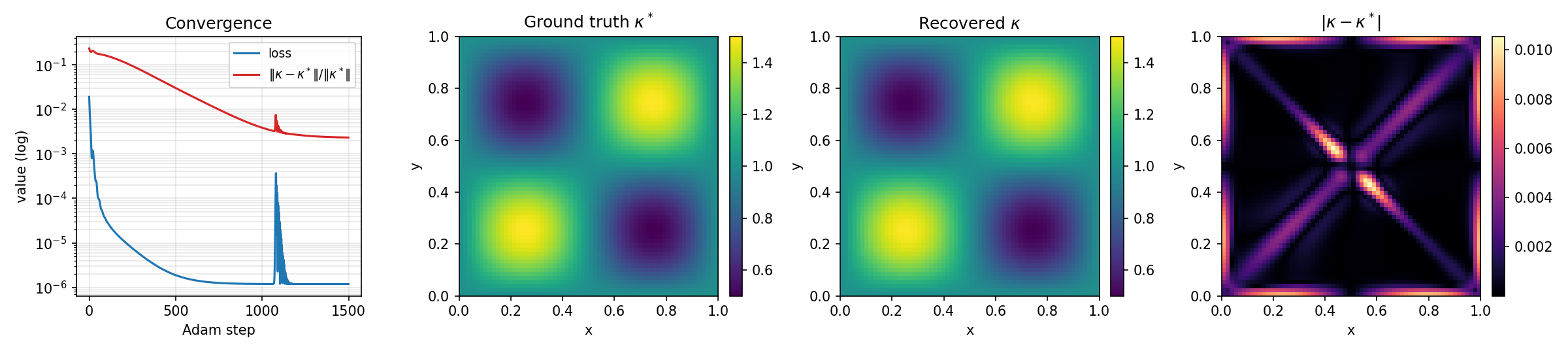}
\caption{Inverse coefficient learning on the variable-coefficient
Poisson equation, $64 \times 64$ grid (3{,}844 unknowns).
\textbf{Left:} loss and relative $L^2$ error
$\|\kappa - \kappa^*\|_2/\|\kappa^*\|_2$ per Adam step (log scale),
both decreasing monotonically over 1500 steps. \textbf{Center:}
ground-truth $\kappa^*$ and recovered $\kappa$ (shared color scale).
\textbf{Right:} pointwise error $|\kappa - \kappa^*|$, below
$1.1 \times 10^{-2}$ everywhere and largest near the boundary and
diagonals where $\nabla u$ is small.}
\label{fig:inverse_problem}
\end{figure*}

\subsection{Gradient Verification}
\label{sec:exp_gradient}

In \S\ref{sec:exp_adjoint}, we verify the linear adjoint gradients against
naive autograd; in this section, we extend the verification to the nonlinear and
eigenvalue cases of \S\ref{sec:adjoint_instances} by comparing
adjoint gradients against centered finite differences,
\begin{equation}
    \frac{\partial \mathcal{L}}{\partial \theta} \;\approx\;
    \frac{\mathcal{L}(\theta + \epsilon) - \mathcal{L}(\theta - \epsilon)}{2\epsilon},
    \quad \epsilon = 10^{-5},
\end{equation}
on randomly perturbed entries of $\bA$ and $\bb$.

\begin{table}[t]
\centering
\caption{Gradient verification for the nonlinear and eigenvalue paths
vs.\ central finite differences (FD); forward/backward cost in units of
forward operations.}
\label{tab:gradient}
\vspace{0.5em}
\scriptsize
\setlength{\tabcolsep}{4pt}
\begin{tabular}{lrll}
\toprule
\textbf{Operation} & \textbf{Rel.\ err.} & \textbf{Fwd} & \textbf{Bwd} \\
\midrule
Eigenvalue ($k{=}6$)\textsuperscript{\dag}     & $2.1 \times 10^{-6}$ & 1 LOBPCG    & outer prod. \\
Nonlinear (5 Newton)     & $4.7 \times 10^{-7}$ & 5 solves    & 1 solve \\
\bottomrule
\end{tabular}
\end{table}

\noindent\textsuperscript{\dag}\,LOBPCG and Lanczos eigensolvers
are exposed by \torchsla{}'s \texttt{.eigsh} entry point on top of
the iterative backends; the autograd wrapper itself is independent
of the eigensolver choice and applies the Hellmann--Feynman gradient
of Eq.~\eqref{eq:adjoint_eig} once at convergence.

Table~\ref{tab:gradient} shows relative errors below $10^{-5}$ for
both paths. The nonlinear 
backward pass runs a single adjoint solve regardless of the five 
Newton iterations performed in the forward; the eigenvalue 
gradient uses Eq.~\eqref{eq:adjoint_eig} directly and requires 
only an $\cO(\text{nnz})$ outer-product evaluation, with no 
additional linear solves.

\subsection{End-to-End: Inverse Coefficient Learning}
\label{sec:exp_inverse}
To demonstrate that gradients flow through \torchsla{} solves 
inside a realistic training loop, we consider an inverse problem 
on the variable-coefficient Poisson equation
\begin{equation}
\begin{aligned}
    -\nabla \cdot \big( \kappa(\bx) \nabla u(\bx) \big)
    &= f(\bx) \quad \text{on } (0,1)^2, \\
    u &= 0 \quad \text{on } \partial\Omega,
\end{aligned}
\end{equation}
with ground-truth conductivity 
$\kappa^*(x,y) = 1 + 0.5 \sin(2\pi x) \sin(2\pi y)$ and 
$f \equiv 1$. We discretize with finite differences on a 
$64 \times 64$ grid, generate observed solutions $u_{\text{obs}}$ 
by a forward solve with $\kappa^*$, and then learn $\kappa$ from 
$u_{\text{obs}}$ alone.

We parametrize $\kappa = \mathrm{softplus}(\theta)$ with
$\theta \in \R^{64\times 64}$ a single \texttt{torch.nn.Parameter}
(softplus enforcing $\kappa > 0$). At every step we assemble the
cell-centered five-point discretization of
$-\nabla\!\cdot\!(\kappa\nabla u) = f$ as a \torchsla{}
\texttt{SparseTensor}, solve $A(\kappa)\, u_{\text{pred}} = f$ via
\texttt{A.solve(f)}, and update $\theta$ with
Adam~\citep{kingma2015adam}
($\mathrm{lr} = 5\times 10^{-2}$) on
$\| u_{\text{pred}} - u_{\text{obs}} \|_2^2$ plus a
Tikhonov~\citep{tikhonov1977solutions}
smoothness regularizer
$10^{-3} \cdot \|\nabla_h \kappa\|_2^2 / N$. Gradients with
respect to $\kappa$ flow through the solve via the adjoint path of
\S\ref{sec:adjoint}, with no custom \texttt{autograd.Function}
written at the user level.

\paragraph{Results.} Figure~\ref{fig:inverse_problem} shows the
optimization trajectory and recovered coefficient field. After 1500
Adam steps (48.6~s on a single NVIDIA RTX PRO 6000 Blackwell,
$\sim$32~ms/step), the recovered
conductivity satisfies
$\|\kappa - \kappa^*\|_2 / \|\kappa^*\|_2 = 2.3 \times 10^{-3}$ and
the corresponding forward solution satisfies
$\|u(\kappa) - u_{\text{obs}}\|_2 / \|u_{\text{obs}}\|_2 = 3.0 \times
10^{-5}$. The recovered field stays in $[0.503, 1.495]$, a tight
match to the ground-truth range $[0.5, 1.5]$. Recovering $\kappa$ to
$0.23\%$ relative error in under a minute confirms that the adjoint
path of \S\ref{sec:adjoint} delivers usable, well-scaled gradients
inside a standard PyTorch \texttt{Adam} loop---the only line of code
specific to differentiable sparse linear algebra is \texttt{A.solve(f)}.

\section{Conclusion}
\label{sec:conclusion}

We presented \torchsla{}, an open-source PyTorch library for
differentiable sparse linear algebra that closes a long-standing
ecosystem gap: a single autograd-aware API spanning direct, iterative,
nonlinear, and eigenvalue solvers across five interchangeable backends
(SciPy, Eigen, cuDSS, CuPy, PyTorch-native), with batched dispatch
through the \texttt{SparseTensor}/\texttt{SparseTensorList} hierarchy
and distributed multi-GPU execution through \texttt{DSparseTensor}. A
unified IFT adjoint keeps the autograd graph at $\cO(1)$ nodes
regardless of iterations or backend, and an autograd-compatible
distributed layer composes domain decomposition with PyTorch autograd
via a transposed halo exchange. On 2D Poisson with H200 GPUs,
\torchsla{} reaches 169M DOF on one GPU and 400M across three, the
adjoint backward gives a $195\times$ memory reduction at 1000 CG
iterations, and an inverse coefficient-learning task recovers a varying
conductivity to $0.23\%$ relative $L^2$ error in under a minute---the
only solver-specific line being \texttt{A.solve(f)}.

\paragraph{Limitations and future work.} The \texttt{pytorch}-native
iterative backend currently supports only Jacobi preconditioning,
insufficient at large DOF---hence the $10^{-2}$ residuals in our
multi-GPU runs under a fixed 1000-iteration budget---and our benchmarks
focus on 2D Poisson; broader validation on 3D PDEs, indefinite
systems, and GNN graph-Laplacians remains future work. The backend
abstraction (\S\ref{sec:backend}) is designed for two extensions:
wrapping mature distributed-sparse libraries
(PETSc~\citep{balay2019petsc}, Trilinos~\citep{trilinos},
hypre~\citep{falgout2002hypre}, AmgX~\citep{naumov2015amgx}) as
backends, and registering \emph{learned} preconditioners (learned
CG~\citep{li2023learning}, graph-neural~\citep{chen2024graph}, and
multigrid-inspired neural solvers~\citep{lyu2026a}) trained end-to-end
against full sparse solves---making \torchsla{} a substrate for
\emph{learnable} sparse solvers at scale, not merely a faster solver.

\newpage
\bibliographystyle{icml2026}
\bibliography{references}

\appendix
\onecolumn

\section{Backend Method Coverage}
\label{app:backends}

This appendix expands the backend abstraction described in
\S\ref{sec:backend}. Table~\ref{tab:backends} lists the concrete
solver families exposed through each backend and the problem regimes
for which the auto-dispatch policy selects them by default.

\begin{table}[h]
\centering
\caption{Backends supported by \torchsla{}. The library auto-selects
a backend by device, problem size, and matrix properties (symmetry
and positive-definiteness are detected automatically); users can
override the choice through a single keyword argument. We list the
principal solver methods exposed by each backend; additional Krylov
variants (e.g.\ GMRES, LGMRES, MINRES, QMR, LSQR) are wrapped where
the underlying library provides them. \torchsla{} does not
reimplement these solvers---it provides a uniform autograd-aware
adapter on top of established backends.}
\label{tab:backends}
\vspace{0.5em}
\small
\begin{tabular}{llll}
\toprule
\textbf{Backend} & \textbf{Device} & \textbf{Solver methods} & \textbf{Recommended for} \\
\midrule
\texttt{scipy}   & CPU  & SuperLU, UMFPACK, CG, BiCGStab & CPU default; machine precision \\
\texttt{eigen}   & CPU  & CG, BiCGStab                   & Alternative CPU iterative \\
\texttt{cudss}   & CUDA & LU, Cholesky, LDLT             & CUDA default; fastest direct ($<\!2$M DOF) \\
\texttt{cupy}    & CUDA & LU, CG, GMRES                  & GPU direct + iterative via \texttt{cupyx} \\
\texttt{pytorch} & CUDA & CG, BiCGStab                   & Very large problems ($>\!2$M DOF) \\
\bottomrule
\end{tabular}
\end{table}

\paragraph{Software versions.} For reproducibility---the auto-dispatch
heuristics and especially the cuDSS API are version-sensitive---we
report the backend versions used for all experiments:
PyTorch~2.10.0 (CUDA~12.8), NVIDIA cuDSS~0.7.1
(\texttt{nvidia-cudss-cu12}), CuPy~14.0.1, SciPy~1.15.3, and
NumPy~2.4.4. Multi-GPU runs use the NCCL backend
bundled with the above PyTorch build.

\paragraph{Single-GPU memory breakdown.} The 443 bytes/DOF reported
for the \texttt{pytorch}-native CG backend in
Table~\ref{tab:benchmark} is $\approx\!3\times$ a minimal-storage
estimate of $\approx\!150$ B/DOF: the COO matrix at $\sim\!5$
non-zeros/row costs 24~B/nnz (an \texttt{int64} row and column index
plus a \texttt{float64} value), and the Jacobi diagonal and the CG
work vectors ($\bx,\br,\bp,\bA\bp$) add a few 8-B/DOF vectors. The
remaining $\sim\!2\times$ is Krylov temporaries, autograd-saved
forward tensors, and CUDA allocator fragmentation.

\section{API Examples}
\label{app:api}

Listing~\ref{lst:api} shows the user-facing API surfaces used
throughout the paper. The examples emphasize that single solves,
backend overrides, batched solves, nonlinear solves, and distributed
solves all preserve the same PyTorch autograd interface: users call a
solver method in the forward pass and receive adjoint gradients through
ordinary \texttt{loss.backward()}.

\begin{lstlisting}[language=Python,caption={\torchsla{} API across single, batched, and distributed solves. Gradients flow through every variant via the same adjoint path.},label={lst:api}]
import torch
from torch_sla import SparseTensor, nonlinear_solve

# 1. Single solve with auto-dispatched backend
val = torch.randn(nnz, requires_grad=True)
A = SparseTensor(val, row, col, shape=(n, n))
b = torch.randn(n, requires_grad=True)
x = A.solve(b)                    # cudss / pytorch / scipy
loss = x.pow(2).sum()
loss.backward()                   # adjoint gradients, O(1) graph

# 2. Explicit backend / method override
x = A.cuda().solve(b.cuda(), backend='cudss', method='cholesky')
x = A.cuda().solve(b.cuda(), backend='pytorch', method='cg')

# 3. Batched solve with shared sparsity pattern
A_batch = SparseTensor(val_batch, row, col, shape=(B, n, n))
x_batch = A_batch.solve(b_batch)  # one symbolic factorization

# 4. Nonlinear solve with adjoint gradients (module-level function)
def residual(u, A_val, f):
    A_local = SparseTensor(A_val, row, col, shape=(n, n))
    return A_local @ u + u**2 - f
u = nonlinear_solve(residual, torch.zeros(n), val, f, method='newton')

# 5. Distributed solve via domain decomposition (multi-GPU)
from torch_sla.distributed import DSparseTensor
A_dist = DSparseTensor.from_global(
    val, row, col, shape, num_partitions=world_size,
    my_partition=rank, device=f'cuda:{rank}')
x_dist = A_dist.solve(b_local, atol=1e-10)
\end{lstlisting}

\section{Distributed Conjugate Gradient and Halo Exchange}
\label{app:dist_cg}

This section gives the operational details behind the distributed
layer of \S\ref{sec:distributed}. Figure~\ref{fig:halo} illustrates
the halo exchange pattern underlying distributed SpMV, while
Algorithm~\ref{alg:dist_cg} shows where halo exchange and global
reductions enter the distributed CG loop. Together they make explicit
which parts of the computation are local to a partition and which parts
require inter-process communication.

\begin{figure}[h]
\centering
\begin{tikzpicture}[
    scale=0.9,
    node/.style={circle, draw, minimum size=7mm, font=\small},
    owned0/.style={node, fill=blue!30},
    owned1/.style={node, fill=red!30},
    halo/.style={node, fill=gray!20, dashed},
    arrow/.style={->, thick, >=stealth}
]
\node[owned0] (n0) at (0,0) {0};
\node[owned0] (n1) at (1.1,0) {1};
\node[owned0] (n2) at (2.2,0) {2};
\node[halo]   (h3) at (3.3,0) {3};
\node[halo]   (h2) at (4.7,0) {2};
\node[owned1] (n3) at (5.8,0) {3};
\node[owned1] (n4) at (6.9,0) {4};
\node[owned1] (n5) at (8.0,0) {5};
\node[above=0.2cm, font=\footnotesize\bfseries] at (1.1,0) {Process 0};
\node[above=0.2cm, font=\footnotesize\bfseries] at (6.35,0) {Process 1};
\draw[thick] (n0) -- (n1) -- (n2) -- (h3);
\draw[thick] (h2) -- (n3) -- (n4) -- (n5);
\draw[arrow, blue!70, bend left=25] (n2) to node[above, font=\scriptsize] {send} (h2);
\draw[arrow, red!70,  bend left=25] (n3) to node[below, font=\scriptsize] {send} (h3);
\draw[dashed, thick, gray] (4.0, -0.7) -- (4.0, 0.9);
\node[gray, font=\scriptsize] at (4.0, -0.95) {boundary};
\node[owned0, scale=0.6] at (0, -1.5) {};
\node[right, font=\scriptsize] at (0.35, -1.5) {Owned P0};
\node[owned1, scale=0.6] at (2.2, -1.5) {};
\node[right, font=\scriptsize] at (2.55, -1.5) {Owned P1};
\node[halo, scale=0.6]   at (4.4, -1.5) {};
\node[right, font=\scriptsize] at (4.75, -1.5) {Halo (ghost)};
\end{tikzpicture}
\caption{Halo exchange in domain decomposition. Each process owns
a subset of nodes (solid colored) and maintains halo copies of
boundary neighbors (dashed). Before each distributed SpMV,
processes exchange updated values at partition boundaries; the
local SpMV then proceeds independently using owned and halo
values.}
\label{fig:halo}
\end{figure}

The same halo-exchange primitive appears once per SpMV inside CG. The
remaining operations are local vector updates or scalar reductions, so
the algorithm follows the standard Krylov recurrence while replacing
each matrix-vector product by a distributed SpMV.

\begin{figure}[h]
\begin{framed}
\refstepcounter{algocounter}
\label{alg:dist_cg}
\small
\noindent\textbf{Algorithm \thealgocounter: Distributed Conjugate Gradient in \torchsla{}.}

\vspace{0.25em}
\noindent\textbf{Input:} distributed $\bA$ (each process holds 
$\bA[\mathcal{O}_p, :]$ and halo metadata), local RHS 
$\bb_{\mathcal{O}_p}$, tolerance $\epsilon$ \\
\textbf{Output:} solution $\bx_{\mathcal{O}_p}$
\begin{enumerate}
    \setlength{\itemsep}{1pt}
    \item $\bx \gets \mathbf{0}$, $\;\br \gets \bb_{\mathcal{O}_p}$, $\;\bp \gets \br$
    \item $\rho \gets \texttt{all\_reduce}(\br^\top\br)$
    \item \textbf{while} $\sqrt{\rho} > \epsilon$ \textbf{do}:
    \begin{enumerate}
        \setlength{\itemsep}{0pt}
        \item $\bA\bp \gets \texttt{DistSpMV}(\bA, \bp)$
              \hfill\textit{// halo exchange + local SpMV}
        \item $\alpha \gets \rho \,/\, \texttt{all\_reduce}(\bp^\top \bA\bp)$
        \item $\bx \gets \bx + \alpha\bp$
        \item $\br \gets \br - \alpha\bA\bp$
        \item $\rho_{\text{new}} \gets \texttt{all\_reduce}(\br^\top\br)$
        \item $\bp \gets \br + (\rho_{\text{new}}/\rho)\,\bp$, 
              $\;\rho \gets \rho_{\text{new}}$
    \end{enumerate}
\end{enumerate}
\end{framed}
\end{figure}

The loop issues two \texttt{all\_reduce} operations per iteration, the
standard form. For very large process counts $P$ the latency of these
two reductions per iteration becomes the bottleneck; pipelined and
communication-avoiding ($s$-step) CG
variants~\citep{hoemmen2010communication} that merge or defer the
inner-product reductions are a natural roadmap item, and compose with
the same transposed-halo backward pass since they only reorganize the
reductions rather than the SpMV.

\section{Adjoint vs.\ Naive: Additional Correctness Analysis}
\label{app:adjoint_extra}

This appendix provides the full sweep behind the memory and timing
summary in \S\ref{sec:exp_adjoint}. Table~\ref{tab:adjoint_vs_naive_full}
reports the measured backward cost at each forced CG iteration count,
and the following paragraph explains the small-problem correctness
check used to compare adjoint gradients against naive autograd after
full convergence.

\begin{table}[h]
\centering
\caption{Adjoint vs.\ naive CG backpropagation, full sweep across 
$k \in \{10, 50, 100, 200, 500, 1000, 2000, 5000\}$.}
\label{tab:adjoint_vs_naive_full}
\vspace{0.5em}
\small
\begin{tabular}{rrrrrr}
\toprule
\textbf{$k$} & \textbf{Adj.\ mem} & \textbf{Naive mem} & \textbf{Adj.\ bwd} & \textbf{Naive bwd} & \textbf{Ratio} \\
\midrule
10    & 320 MB & 0.97 GB  & 35 ms  & 57 ms    & 3.0$\times$ \\
50    & 328 MB & 3.52 GB  & 9 ms   & 53 ms    & 11$\times$ \\
100   & 328 MB & 6.71 GB  & 13 ms  & 104 ms   & 20$\times$ \\
200   & 328 MB & 13.1 GB  & 25 ms  & 204 ms   & 41$\times$ \\
500   & 328 MB & 32.2 GB  & 60 ms  & 508 ms   & 100$\times$ \\
1000  & 328 MB & 64.1 GB  & 117 ms & 1015 ms  & 195$\times$ \\
2000  & 328 MB & OOM      & 230 ms & ---      & --- \\
5000  & 328 MB & OOM      & 332 ms & ---      & --- \\
\bottomrule
\end{tabular}
\end{table}

On a smaller problem ($n_{\text{grid}} = 64$, $N = 4096$), where both 
paths can be run to full convergence (atol $= 10^{-12}$, 
$k = 3000$), the loss values agree to machine precision 
($1.96 \times 10^{-16}$ relative error), the gradient 
$\partial \mathcal{L}/\partial \bb$ matches to 
$2.6 \times 10^{-14}$, and the matrix gradient 
$\partial \mathcal{L}/\partial \bA$ matches to 
$6.8 \times 10^{-4}$ relative error. The looser agreement on the 
matrix gradient reflects floating-point round-off accumulated by 
naive backpropagation over 3000 iterations of CG recurrences, whereas 
the adjoint path evaluates the closed-form sparse outer product 
$-\lambda_i x_j$ once at the converged solution. This explains why the 
adjoint method remains numerically robust at high iteration counts,
in addition to reducing memory and time.

\section{Extended Comparison with Related Libraries}
\label{app:related}

Table~\ref{tab:jax_vs_torch} and \S\ref{sec:related} place \torchsla{}
among differentiable linear-algebra libraries; here we expand the two
largely orthogonal axes that most cleanly separate them.

\paragraph{Operator representation: explicit vs.\ implicit.}
\torchsla{} stores an \emph{explicit} sparse matrix
(\texttt{val}/\texttt{row}/\texttt{col}), which is exactly what heavy
sparse \emph{direct} factorizations (cuDSS LU/Cholesky/LDLT) and
domain-decomposition partitioners consume. CoLA instead composes
\emph{implicit}, matrix-free \texttt{LinearOperator}s: it can factor
$\bA\otimes\bB$ and apply matrix functions $f(\bA)$ efficiently, but it
does not expose the non-zero pattern a sparse direct solver or an
ILU/AMG preconditioner needs. Consequently CoLA reaches \texttt{logdet},
\texttt{trace}, and $f(\bA)$ on structured operators that \torchsla{}
does not target, while \torchsla{} reaches GPU sparse-direct
factorization and tensor-parallel distribution that CoLA does not.

\paragraph{Structure exploited: sparsity vs.\ algebraic composition.}
\torchsla{} and PhiFlow exploit the \emph{sparsity pattern}; CoLA
exploits \emph{algebraic} composition (Kronecker, block, low-rank,
FFT). The two are orthogonal, and the choice dictates the admissible
preconditioners: a pure-matvec black box admits only Jacobi/polynomial
preconditioners, whereas ILU/IC/AMG need the explicit non-zeros. This
is why PhiFlow materializes its operators (via
\texttt{jit\_compile\_linear}) precisely to enable ILU, and why
\torchsla{}'s explicit representation makes the AmgX/PETSc backends on
its roadmap natural.

\paragraph{Backend reach.} Theseus's sparse solvers (CHOLMOD, cudaLU,
BaSpaCho) are internal to its nonlinear-least-squares optimizer rather
than a general-purpose API. PhiFlow's Krylov solvers are
backend-agnostic (JAX/PyTorch/TF) but offer no GPU sparse-direct path
(its direct solve routes to CPU SciPy/SuperLU). In JAX, the capability
\torchsla{} exposes in one library corresponds to the
Lineax\,+\,JAXopt\,+\,Optimistix trio together; \torchsla{}'s
contribution is to unify these axes in PyTorch with a GPU
sparse-direct backend and a tensor-parallel distributed layer.

\end{document}